\documentclass[a4paper, superscriptaddress, nofootinbib, amsmath, amssymb, amsfonts, aps, prapplied, showkeys, longbibliography, reprint]{revtex4-2}

\usepackage[utf8]{inputenc}

\usepackage{printlen}
\usepackage{adjustbox}
\usepackage{placeins}

\usepackage{graphicx}
\usepackage{dcolumn}
\usepackage{bm}
\usepackage{hyperref}
\usepackage[mathlines]{lineno}

\usepackage{epstopdf}
\usepackage[T1]{fontenc}
\usepackage{url}
\usepackage{ulem}
\usepackage{changes}
\usepackage{soul}
\usepackage{float}

\usepackage[english]{babel}
\begin{document}

\preprint{APS/123-QED}

\title{Optimization of the degenerate optical parametric oscillations threshold in bichromatically pumped microresonator}


\author{Nadezhda S. Tatarinova}  
\affiliation{Russian Quantum Center, Skolkovo, Moscow 121205, Russia}
\affiliation{Moscow Institute of Physics and Technology, Dolgoprudny, Moscow Region 141701, Russia}
\author{Artem E. Shitikov} 
\affiliation{Russian Quantum Center, Skolkovo, Moscow 121205, Russia}
\author{Georgy V. Grechko}  
\affiliation{Moscow Institute of Physics and Technology, Dolgoprudny, Moscow Region 141701, Russia}
\author{Alexander K. Vorobyev}  
\affiliation{Russian Quantum Center, Skolkovo, Moscow 121205, Russia}
\affiliation{Moscow Institute of Physics and Technology, Dolgoprudny, Moscow Region 141701, Russia}
\author{Anatoly~V.~Masalov}
\affiliation{Russian Quantum Center, Skolkovo, Moscow 121205, Russia}
\affiliation{Lebedev Physical Institute, Russian Academy of Sciences, 119991 Moscow, Russia}%
\author{Igor A. Bilenko}
\affiliation{Russian Quantum Center, Skolkovo, Moscow 121205, Russia}
\affiliation{Faculty of Physics, Lomonosov Moscow State University, 119991 Moscow, Russia}
\author{Dmitry A. Chermoshentsev}
\email{d.chermoshentsev@gmail.com}
\affiliation{Russian Quantum Center, Skolkovo, Moscow 121205, Russia}
\affiliation{Moscow Institute of Physics and Technology, Dolgoprudny, Moscow Region 141701, Russia}
\author{Valery E. Lobanov}
\email{vallobanov@gmail.com}
\affiliation{Russian Quantum Center, Skolkovo, Moscow 121205, Russia}

\date{\today}
\begin{abstract}

Integrated microring resonators have a broad range of applications in diverse fields with the potential
to design compact, robust, energy-efficient devices crucial for quantum applications. Degenerate optical
parametric oscillations (DOPOs) realized in dual-pumped microring resonator with third-order optical
nonlinearity are of special interest. They demonstrate both bistability of the phase of the excited signal
mode and generation of nonclassical light, which can be used for coherent photonic computing. Using cou-
pled mode equations, we perform a comprehensive numerical analysis of DOPO conditions with normal
group velocity dispersion and with bichromatic pumping. Through analytical and numerical approaches,
we identify optimal setup parameter values that minimize the threshold power, highlighting the importance
of considering the full spectrum of mode interactions. Additionally, we show that dispersion engineering,
achievable in photonic molecules or photonic crystal microresonators, may provide a targeted frequency
shift of specific microresonator modes resulting in pump power threshold reduction.
\end{abstract}

\maketitle

\section{\label{sec:intro}Introduction}

Integrated photonic devices provide on-chip, robust, and scalable solutions for a wide range of applications, spanning from fundamental science to industrial technologies \cite{vahala2003optical, Marpaung2013Integrated, lin2018midinfrared, Cheben2018Subwavelength, Perez2020Principles, Zhu2021Integrated}. These applications include laser physics, spectroscopy, metrology, microwave photonics, quantum computing, communications, and sensing \cite{Pasquazi2012Stable, Liang2016Integrated, Weimann2017Silicon, marin2017microresonator, heylman2017optical, Yang2021squeezed, Pelucchi2022potential, Zhang2024Integrated, Labonte2024Integrated}. Advancements in fabrication techniques have enabled the production of CMOS-compatible, on-chip integrated waveguides and microring resonators (MRRs) with low optical loss and precise dispersion control, making them highly suitable for realizing efficient nonlinear and quantum optical interactions \cite{Herr2012universal, Herr2014Temporal, strekalov2016nonlinear, kippenberg2018dissipative}. These developments have laid the foundation for a new generation of compact MRR-based devices, such as frequency comb generators \cite{Kippenberg2011Microresonator, wang2016dual}, microwave generators \cite{kudelin2024photonic, li2025universal}, ultrastable lasers \cite{guo2022chip, kondratiev2023recent}, sources of non-classical states of light \cite{dutt2015onchip, Guo2019Nonclassical, Vaidya2020Broadband}, and random number generators \cite{okawachi2016quantum}.

Among the nonlinear effects in microresonators, the excitation of degenerate optical parametric oscillations (DOPO) is of particular interest due to its applications in coherent computing, quantum computing, and random number generation \cite{Marandi2012alloptical, wang2013coherent, Inagaki2016Largescale,  okawachi2016quantum}. Experimental realizations of DOPO have been demonstrated in media exhibiting either second-order $\chi^{(2)}$ or third-order $\chi^{(3)}$ nonlinearities \cite{Marandi2014Network, yamamoto2017coherent, vernon2019scalable, Bruch2019Onchip, okawachi2020demonstration}.

One of the techniques for DOPO realization in MRRs with $\chi^{(3)}$ nonlinearity is based on the degenerate four-wave mixing process and involves bichromatic (or dual) pumping of the microresonator modes \cite{Hansson2014Bichromatically}. Both theoretical and experimental studies have shown that MRRs with normal group velocity dispersion (GVD) are best suited for DOPO realization in dual-pumped setups \cite{okawachi2015dual}. The generated signal exhibits bistability between two phase states shifted by $\pi$, a crucial feature for implementing all-optical coherent Ising machines based on photonic integrated circuits \cite{Tezak2020Integrated, okawachi2020demonstration, Mohseni2022Ising, Li2024Scalable}. Furthermore, below the DOPO pump power threshold, the MRR can function as a generator of squeezed states of light \cite{zhao2020neardegenerate}. However, fine-tuning of MRR properties, which could further optimize DOPO performance, remains an open area of research.

It is known that dispersion engineering significantly affects nonlinear processes in microresonators. 
Namely,  the selective shifting of mode eigenfrequencies can enhance or suppress certain nonlinear processes, including parametric oscillation, the generation of frequency combs, and dissipative solitons \cite{fujii2018analysis, Yu2021Spontaneous, Black2022optical, seifoory2022degenerate, helgason2023surpassing, ulanov2025quadrature}.
Though the width and height of the waveguide are the main fabrication parameters that determine the dispersion profile in integrated microresonators,
they are not the only factors \cite{pfeiffer2016photonic, pfeiffer2018photonic, Pfeiffer2018Ultra,  fujii2020dispersion, Anderson2022Zero}. For instance, a prominent dispersion engineering technique exploits avoided mode crossings in coupled resonator systems, also known as photonic molecules \cite{Xue2019Superefficient, helgason2021dissipative, helgason2023surpassing}. Another approach involves periodically varying the ring width, which alters the effective refractive index, forming a photonic crystal structure and leading to targeted mode splitting \cite{lu2014selective, moille2023fourier, lucas2023tailoring}. Such structures offer intriguing possibilities for compact chip-based devices \cite{Yu2021Spontaneous, Yu2022continuum, Lu2022high, Black2022optical, Yang2022Multidimensional, moille2023fourier, lucas2023tailoring, lobanov2024generation, Lu2024band, Spektor2024Photonic, ulanov2024synthetic, Zang2025laser}. We also applied targeted mode shifting to investigate its influence on the DOPO threshold. A comprehensive study on key parameters, including the second- and higher-order dispersion, frequency interval between the pumps, and their impact on the efficiency of dual-pumped MRR-based DOPOs, is still missing. 

In this work, we theoretically investigate how parameters of the MRR with Kerr nonlinearity and the dual pumps affect the parametric threshold power and explore the methods to decrease it. We start with a simplified three-mode model, which reveals a minimum in the DOPO threshold in terms of pump power. Next, we perform a numerical study using a multimode model, in which we systematically analyze the dependence of the threshold pump power on the modeled system's parameters. While the optimal parameters are also found, they differ significantly from those of the three-mode model due to the influence of additional nonlinear frequency conversion processes. Finally, we numerically demonstrate that controlled dispersion engineering---specifically, a shift of the signal mode and a symmetrical shift of the particular sidebands---can substantially reduce the DOPO generation threshold. This approach paves the way for the new generation of energy-efficient optical sources for different quantum applications.

\section{\label{sec:methods}Methods}

\subsection{\label{ssec:optimalparams}Coupled mode equations}

To model field dynamics inside the MRR with the bichromatic pumping, we use a set of normalized coupled mode equations (CMEs) \cite{chembo2010modal, Herr2012universal}:
\begin{multline}\label{eq:CME}
    \frac{\partial a_{\mu}}{\partial\tau} = -\left[1 - i \zeta_{\mu} \right] a_\mu + \delta_{+n,\mu} f_{+n} + \delta_{-n,\mu} f_{-n} + \\
    +i \sum_{\mu' \leqslant \mu''} \left( 2 - \delta_{\mu',\mu''}\right) a_{\mu'} a_{\mu''}a_{\mu' + \mu'' - \mu} ^*.
\end{multline}
Here $a_\mu$ is the normalized complex amplitude of the $\mu$-th~mode corresponding to the cold MRR frequency $\omega_\mu$, with relative eigenmode number $\mu$. The dispersion of the cold MRR resonant frequencies is described by:
\begin{equation}\label{eq:dispersion}
    \omega_\mu = \omega_0 + D_1\mu + \frac{1}{2} D_2 \mu^2 + \frac{1}{6}D_3\mu^3 + \dots,
\end{equation}
where $D_1/ 2\pi$ is the MRR's free spectral range (FSR), $D_2/ 2\pi$ is the second-order dispersion coefficient and $D_3/ 2\pi$ is the third-order dispersion coefficient. The case $\mu = 0$ corresponds to the mode $\omega_0$, which is nearest to the arithmetic mean of the pump frequencies $\omega_{p\pm}$. $\tau = \kappa t / 2$ denotes the normalized slow time, where $\kappa$ is the cavity decay rate, consisting of both internal and coupling losses; $\delta_{\mu', \mu ''}$ is the Kronecker delta. The normalized pump amplitudes are given by $f_{\pm n} = \sqrt{\frac{8 \eta \hbar \omega_{0}^2 c n_2}{\kappa^2n_0^2 V_\text{eff}}} \sqrt{\frac{P_{\text{in}, \pm}}{\hbar \omega_{0}}}$, where $\pm n$ is the relative eigenmode number of the MRR closest to the corresponding pump frequency, $\eta$ is the ratio of coupling losses to the total cavity decay rate, $P_{\text{in}, \pm}$ are the input pump powers, $n_0$ is the refractive index, $n_2$ is the nonlinear refractive index, $V_\text{eff}$ is the effective mode volume, $c$ is the speed of light in vacuum, $\hbar$ is the reduced Planck constant.

The terms $\zeta_\mu$ represent the normalized frequency detunings from the MRR's equidistant frequency grid:
\begin{equation}\label{eq:norm_det}
    \zeta_\mu = \frac{\zeta_{+n} + \zeta_{-n}}{2} + \frac{\zeta_{+n} - \zeta_{-n}}{2n}\mu + \frac{d_2}{2} (n^2 - \mu^2),
\end{equation}
with the normalized pump frequency detunings $\zeta_{\pm n } = 2\left({\omega_{p\pm} - \omega_{\pm n }}\right)/ {\kappa}$ and $d_2 = 2{D_2}/{\kappa}$ is the normalized second-order dispersion coefficient. We limit the dispersion law to the second-order approximation for primary study. 
The use of normalized units allows us to disregard specific material properties of the MRR, including the quality factor and nonlinear strength, thereby enhancing the generality of the analysis and making the results applicable to a wide range of photonic systems.
More details regarding the normalization of these equations can be found in Appendix~\ref{ap:CMEs}.

From Eq.~(\ref{eq:CME}) and Eq.~(\ref{eq:norm_det}) it is evident that the system dynamics is fully characterized by the normalized pump parameters ($\pm n, f_{\pm n}, \zeta_{\pm n}$) and a single normalized MRR parameter ($d_2$). Our objective is to determine the optimal parameters of the dual-pumped MRR system for DOPO realization, focusing on minimizing power consumption while ensuring that the signal mode $a_0$ is excited. Since we are analyzing a dual-pumped system, the total pump power $|f_{+n}|^2 + |f_{-n}|^2$ is chosen as one of the parameters for optimization.

\subsection{\label{ssec:TMM}Analytical three-mode model}

\begin{figure*}[ht]
    \centering
    \includegraphics[width=1\linewidth]{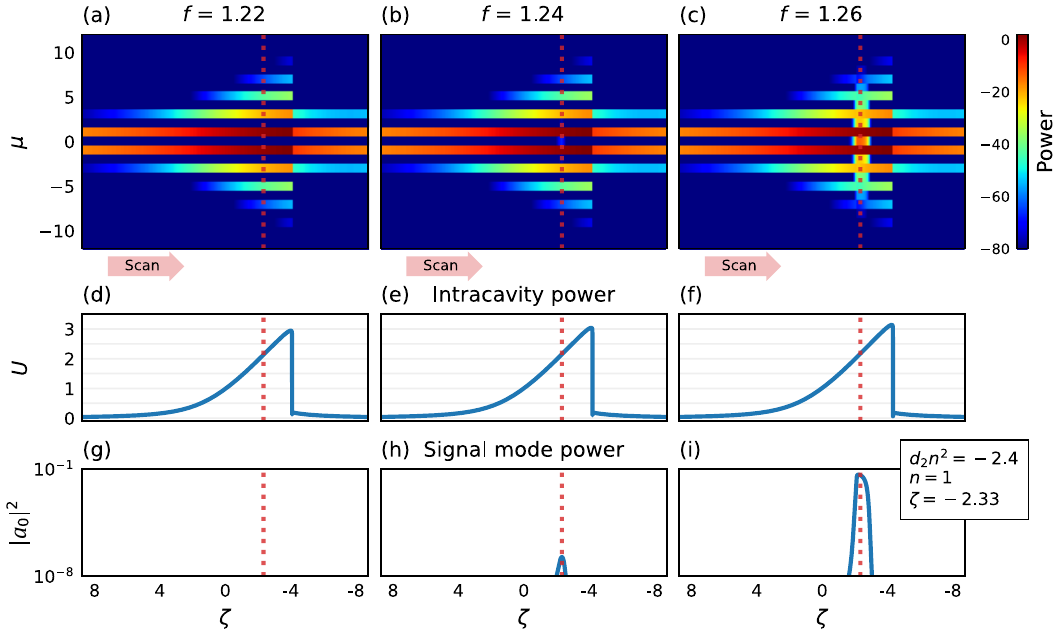}
    \caption{Numerical simulation results for different pump amplitude values $f$ with $d_2n^2 = -2.4$ and $n = 1$.  (a-c) Spectrograms for different pump regimes: power below the threshold (a), at the threshold (b), and above the threshold (c). 
    Colorbar indicates spectrum components’ relative power in a logarithmic scale: $10\lg|a_\mu|^2$. (d-f) Intracavity power $U = \sum_\mu |a_\mu|^2$ in all regimes, respectively. (g-i) Signal mode power $|a_0|^2$ in all regimes, respectively. Red dashed lines denote the detuning $\zeta=-2.33$ corresponding to the generation of the signal mode $a_0$, $\mu = 0$ in the threshold regime (h). All quantities are plotted in dimensionless units.}
    \label{fig:fig1}
\end{figure*}

We initiate our analysis with a simplified scenario involving only three interacting modes: two pumped modes with indices $\pm n$ (denoted as $a_{+n}$ and $a_{-n}$ for modes $+n$ and $-n$, correspondingly) and the signal (central) mode $a_0$ with index $\mu = 0$. For the near-the-threshold regime, we assume $|a_0|^2 \ll |a_{\pm n}|^2$.

The dynamics of the pumped modes can be described by the following equations:
\begin{multline}\label{eq:tmm_pumps}
    \frac{\partial a_{\pm n}}{\partial \tau} = -\left[1 - i \zeta_{\pm n} \right]a_{\pm n}+ \\ 
    + i \left(|a_{\pm n}|^2 + 2 |a_{\mp n}|^2\right)a_{\pm n} +  f_{\pm n}. 
\end{multline}
Assuming a steady-state solution ($\partial a_{\pm n} / \partial \tau = 0$) for mode amplitudes $a_{\pm n}$, we obtain:
\begin{equation}
    \frac{|f_{\pm n}|^2}{|a_{\pm n}|^2} = 1 + \left( \zeta_{\pm n} + |a_{\pm n}|^2 + 2 |a_{\mp n}|^2 \right)^2, \label{eq:tmm_st_state}
\end{equation}
and the minimal total pump power is achieved with detunings
\begin{equation}
    \zeta_{\pm n}  = - |a_{\pm n}|^2 - 2 |a_{\mp n}|^2, \label{eq:dets_pm}
\end{equation}
and equals
\begin{equation}\label{eq:totalpow}
    |f_{+n}|^2 + |f_{-n}|^2 = |a_{+n}|^2 + |a_{-n}|^2.
\end{equation}
So, minimizing total pump power is equivalent to minimizing $|a_{+n}|^2 + |a_{-n}|^2$.

Equation for the signal mode is given by:
\begin{multline}\label{eq:tmm_signal}
    \frac{\partial a_{0}}{\partial\tau} = -\left[1 - i \zeta_{0} \right] a_0 +  i \left(2|a_{+n}|^2
    + 2 |a_{-n}|^2\right)a_0 + \\
    +2i a_{+n}a_{-n}a_0^*, 
\end{multline}
where 
\begin{multline}\label{eq:det_zero}
        \zeta_0 = \frac{\zeta_{+n} + \zeta_{-n}}{2} + \frac{d_2}{2}n^2 =\\ 
        =-\frac{3}{2} \left( |a_{+n}|^2 + |a_{-n}|^2\right) + \frac{d_2}{2}n^2. 
\end{multline}
This equation can be divided into separate equations for real and imaginary parts of $a_0$ (see Ref.~\cite{okawachi2020demonstration}). The eigenvalues of the coefficient matrix for this system are:
\begin{equation}\label{eq:lam_eq}
    \lambda_{1,2}= -1 \pm \sqrt{4\left| a_{+n}a_{-n}\right|^2 - \left[\zeta_0 + 2\left(|a_{+n}|^2 + |a_{-n}|^2\right)\right]^2}.    
\end{equation}
The signal mode can be excited only if $\text{Re}~\lambda > 0$. At the threshold, where $\lambda = 0$, the following relationship is obtained:
\begin{equation}\label{eq:lam_thr}
    \sqrt{4\left| a_{+n}+a_{-n}\right|^2 - \frac{1}{4}\left[d_2n^2 + \left(|a_{+n}|^2 + |a_{-n}|^2\right)\right]^2} = 1.
\end{equation}
By solving this equation for $|a_-|^2$ and  subsequently minimizing the function $|a_+|^2 + |a_-|^2$, it can be determined that the minimum is achieved under the equal mode amplitude condition $|a_{+n}|^2 = |a_{-n}|^2,$ which leads [from Eq.~(\ref{eq:dets_pm})] to the necessary condition of equal detunings $\zeta_{+n} = \zeta_{-n}$ and pump powers $|f_{+n}|^2 = |f_{-n}|^2$. The simplest way to satisfy these conditions is to use the symmetric frequency scan. Taking this into consideration, we also derive the threshold power of each pump $f_\text{thr} = |f_{+n}| = |f_{-n}|$:
\begin{equation}\label{eq:tmm_fthr_curve_eq}
    |f_\text{thr}|^2 = \frac{1}{6} \left(d_2 n^2 + 2 \sqrt{3 + \left(d_2 n^2\right)^2}\right).
\end{equation}
This expression indicates that the threshold power depends on $d_2 n^2$ parameter.
The minimum is reached for $d_2 n^2 = -1$, indicating that normal GVD is required.

From this simplified three-mode model, we obtain analytically the optimal parameters of the MRR and pumps that minimize the input pumping power:
\begin{equation}\label{eq:tmm_min_params}
    |f_{+n}|^2 = |f_{-n}|^2 = \frac{1}{2},~\zeta_{+n} = \zeta_{-n}= -\frac{3}{2},~d_2n^2 = -1.
\end{equation}

\begin{figure}[tp]
    \centering
    \includegraphics[width=\linewidth]{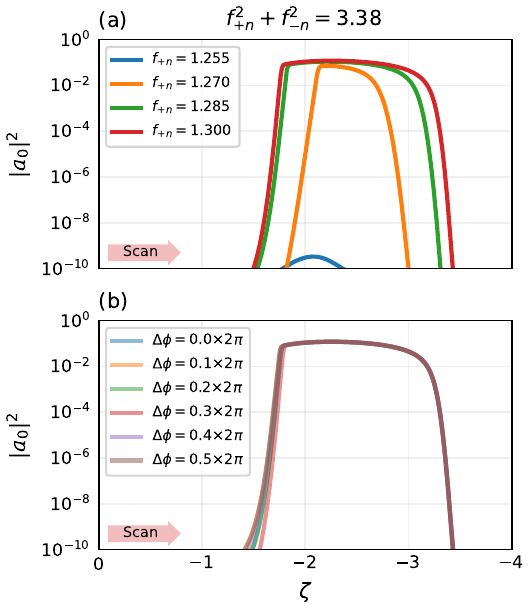}
    \caption{
    Signal mode power $|a_0|^2$ under pump asymmetry conditions. (a) Total pump power fixed at $f_{+n}^2 + f_{-n}^2 = 3.38$, corresponding to the symmetric case where $f = 1.3$, i.e., $2f^2 = 3.38$. (b) Phase mismatch $\Delta\phi$ between pumps with fixed amplitudes $f =1.3$. All simulations use $d_2n^2 = -2.4$ and $n=5$. All quantities are plotted in dimensionless units.
    }
    \label{fig:fig8}
\end{figure}

\subsection{\label{ssec:num}Numerical simulation}

The examined three-mode model does not account for additional nonlinear processes of the multimode interaction present in real-life scenarios, requiring additional analysis \cite{stone2022conversion}. Therefore, we numerically solve a system of 256 CMEs for the dual-pumped MRR [Eq.~(\ref{eq:CME})] using Fast Fourier Transform methods for nonlinear terms calculation \cite{hansson2012numerical}. We verified that a further increase in the number of CMEs does not change the results. The CME approach allows us to observe the dynamics of individual modes directly.
We use a symmetrical linear frequency scan ($\zeta = \zeta_{+n} = \zeta_{-n}, f = f_{+n} = f_{-n}$), moving from the blue-detuned region to the red-detuned by progressively decreasing $\zeta$, for different $d_2, n, f$ parameters.
The higher-order dispersion coefficients are neglected. To account for vacuum fluctuations in this system, we initialize each eigenmode of the MRR with a noise-like seed input having a mean amplitude of $10^{-4}$ at every step of the simulation. 

The threshold power is determined by monitoring the intracavity power of the signal mode and identifying the point at which it exceeds the initial noise level. Notably, reducing the scan rate does not influence the threshold.
The simulation results for different pump powers and $d_2n^2=-2.4$ are presented in Fig.~\ref{fig:fig1}. For $f$ below the threshold, only the non-degenerate processes occur, forming a few teeth of the primary comb with $2n\times$FSR spacing [Fig.~\ref{fig:fig1}(a)]. In the threshold regime, the signal mode $\mu = 0$ is generated at the particular detuning $\zeta = -2.33$ [Fig.~\ref{fig:fig1}(h)]. For the further increasing pump power, the signal mode power is increasing and the lines between primary comb teeth start generating [Fig.~\ref{fig:fig1}(c)]. Moreover, we verified that the proposed threshold condition accurately identifies the point at which parametric oscillations occur. We observe the quadrature-squeezed vacuum if the pump power is below the threshold, and phase-bistable signal if the power exceeds this threshold (see Appendix~\ref{ap:squeez}).

\begin{figure}[tp!]
    \centering
    \includegraphics[width=\linewidth]{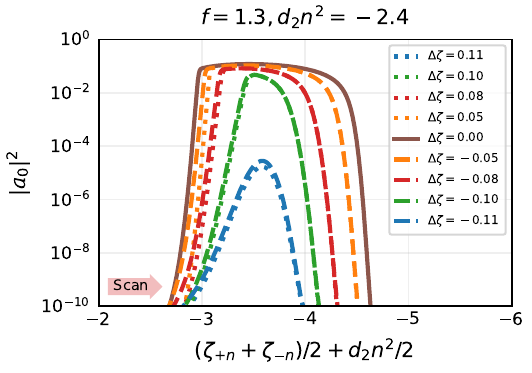}
    \caption{
    Signal mode power $|a_0|^2$ under pump asymmetry conditions with detunings offset $\zeta_{+n} - \zeta_{-n} = \Delta \zeta$. Dotted lines correspond to positive detunings mismatch $\Delta \zeta > 0$ and dashed lines correspond to $\Delta \zeta < 0$. Solid line denotes symmetric scan $\zeta_{+n} = \zeta_{-n}.$ All lines are plotted against the effective detuning of the central mode $\zeta_0$. All simulations use $f=1.3$, $d_2n^2 = -2.4$ and $n=5$. All quantities are plotted in dimensionless units.
    }
    \label{fig:fig9}
\end{figure}
 
The three-mode model predicts the minimal threshold power under the condition of equal pump detunings and powers. A symmetric linear frequency scan offers a direct and practical approach to fulfill these requirements.
However, ideal symmetry can be challenging to achieve experimentally due to noise and system instabilities.
We briefly analyze how deviations from pump symmetry---namely power, detuning, and phase mismatches---affect parametric signal generation. All simulations are conducted in the above-threshold regime, where the DOPO generation region is broad and well-defined.

We fix the total pump power as $f_{+n}^2 + f_{-n}^2 = 3.38$, corresponding to the symmetric case $f_{+n} = f_{-n} = 1.3$, and set $d_2n^2 = -2.4$. The resulting central mode power is shown in Fig.~\ref{fig:fig8}(a). As the amplitude difference between the two pumps increases, the range of detunings supporting DOPO generation narrows. However, a pump amplitude mismatch of up to 2\% does not significantly affect the generation region, indicating moderate tolerance to asymmetry.

We also explore the impact of relative phase offsets between the two pumps. Assuming equal amplitudes $|f_{+n}| = |f_{-n}| = 1.3$, we vary the phase difference $ \Delta\phi$. The results in Fig.~\ref{fig:fig8}(b) reveal no impact on central mode power in the above-threshold regime, indicating that the system is insensitive to pump phase mismatches in this regime.

To study the effect of detuning asymmetry, we fix the detuning of one pump and vary the other. The power of the central mode ($ \mu = 0 $) is plotted as a function of its effective detuning from Eq.~(\ref{eq:det_zero}) in Fig.~\ref{fig:fig9}, corresponding to both positive and negative detuning offsets $\Delta \zeta  =\zeta_{+n} - \zeta_{-n} $. For small mismatches $|\Delta \zeta| < 0.05$, no noticeable change is observed. With increasing mismatch, the range over which parametric generation occurs becomes noticeably narrower.
As the pump power increases, further above the threshold, the parametric generation region widens, leading to greater robustness against pump asymmetry.

Thus, we confirm that the threshold parameters obtained via symmetric frequency scan correspond to the minimal achievable threshold power, as all tested asymmetric configurations result in higher values and shrink the parametric generation region.
Moreover, the system demonstrates resilience to moderate mismatches in the pump detunings and powers, and is insensitive to the phase offset, particularly above the threshold. The required precision is experimentally feasible, and symmetric scans in dual-pump configurations have been demonstrated in \cite{ulanov2025quadrature, zhang2021squeezed}.

\begin{figure}[ht]
    \centering
    \includegraphics[width=\linewidth]{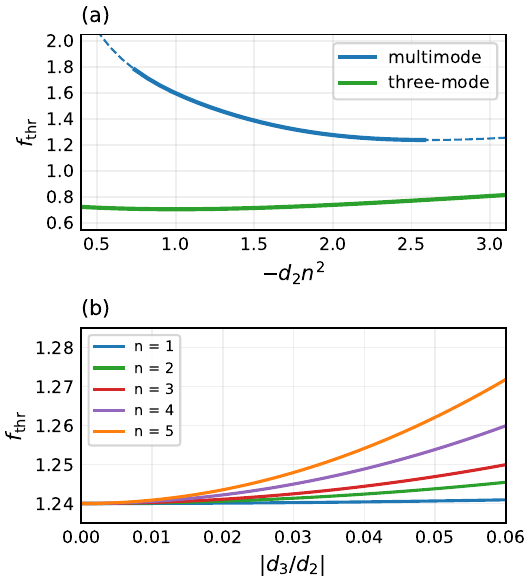}
    \caption{(a) Threshold  amplitude $f_\text{thr}$ dependence on the dispersion parameter $-d_2n^2$ for the analytical solution of the three-mode model (green line) and the numerical solution of the multimode model (blue line). The dashed line shows the areas where the other modes between the pumps in case of $n>1$ are already excited. (b) Threshold amplitude as a function of the added third-order dispersion for a fixed $d_2n^2 = -2.4$ parameter for different indices $n$ of pumped modes. All quantities are plotted in dimensionless units.}
    \label{fig:fig2}
\end{figure}

\begin{figure*}[ht]
    \centering
    \includegraphics[width=1\linewidth]{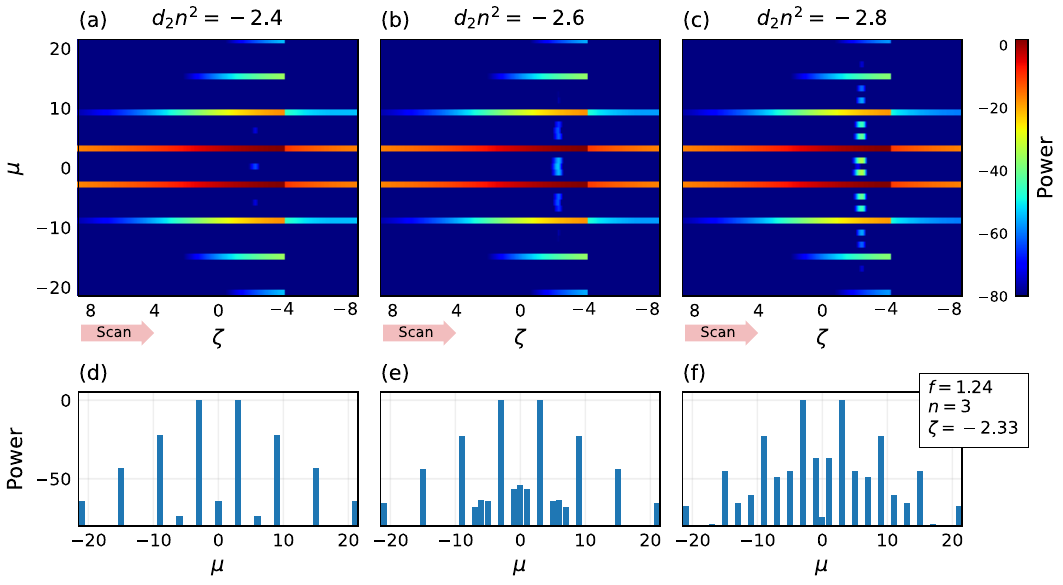}
    \caption{Spectrograms (a-c) and spectra for pumps detuning $\zeta=-2.33$ (d-f) for different dispersion values for fixed near-threshold pump amplitude $f=1.24$. For lower absolute dispersion the signal mode starts generating earlier (a, d), for higher absolute dispersion values the adjacent modes appear earlier  despite the generation of the signal mode (c, f). In all cases the signal mode $\mu = 0$ is excited. Threshold values were calculated for $n = 3$. All quantities are plotted in dimensionless units.}
    \label{fig:fig3}
\end{figure*}

The threshold power values found for the wide range of dispersion parameters are presented in Fig.~\ref{fig:fig2}(a). The green line corresponds to the analytical three-mode model, and the blue line is the multimode approach results. The threshold power value shows a significant increase from its analytical counterpart, with the new minimal pump amplitude $f_\text{thr} = 1.24$  corresponding to a dispersion parameter near $d_2 n^2 = -2.6$. It is noteworthy that, similar to the three-mode model, the threshold power remains independent of the indices of pumped modes when the $-d_2n^2$ parameter is fixed. The dashed areas along the blue line correspond to the threshold of the parametric signal for $n=1$. For the case $n>1$ other modes between the pumps are already excited prior to the signal one. Fig. \ref{fig:fig3} shows the spectrograms and intracavity spectra at the detuning $\zeta=-2.33$, corresponding to the central mode generation, at different values of $d_2 n^2$ and with the 6-FSR interval between pumps. One may see that with the increasing of $|d_2| n^2$ for the fixed pump amplitude $f = 1.24$ the power of the modes with indices $\mu = \pm 1$ becomes pronounced instead of the signal mode $\mu = 0$. One should take into account this fact in the experiment. 

From the numerical study of the multimode model we obtain optimal parameters of the pump and microresonator $|f_\text{thr}| = 1.24,~d_2n^2 = -2.6$ that differ significantly from the analytical ones in Eq.~(\ref{eq:tmm_min_params}) due to the nonlinear interactions between all modes of the MRR. At low $|d_2|n^2 < 0.7$ the threshold power rapidly increases. The value of $d_2n^2$ determines the relatively high value of the frequency interval between pumps $n$, since $d_2$ value is usually of order of magnitude $10^{-1}-10^{-2}$. On the other hand, with the higher $n$ the influence of the higher-order dispersion becomes  more pronounced.

To study the influence of higher-order dispersion on the threshold power values, we extend the normalized dispersion equation derived from Eq.~(\ref{eq:norm_det}):
\begin{equation}
    \zeta_\mu^\text{third-order} = \zeta_\mu +  \frac{d_3}{6} \left(\mu n^2 - \mu^3\right),
\end{equation}
where $d_3 = 2D_3/\kappa$ is the normalized third-order dispersion coefficient. In our modeling, we introduce a small deviation from the second-order approximation using the ratio $|d_3/d_2|$ with the parameter $d_2n^2 = -2.4$ fixed. The outcome shows that increasing $n$ results in a rapid increase of the threshold power [Fig.~\ref{fig:fig2}(b)]. However, up to $|d_3/d_2| < 0.01$ the threshold weakly depends on $n$ and for the further study we assume $d_3 = 0$.

\subsection{\label{ssec:shift}Performance enhancement via dispersion engineering}

Innovative fabrication techniques provide an opportunity to manipulate the dispersion law of MRR precisely. One of the methods proven to enhance nonlinear processes is the targeted mode shifting \cite{Xue2015Mode, lobanov2020generation, Yang2022Stability, helgason2023surpassing, Lobanov2023Platicon}, which can be experimentally realized using photonic molecules (system of coupled microresonators) \cite{helgason2021dissipative, zhang2021squeezed, rebolledo2023platicon, Sanyal2025Nonlinear} or photonic crystals \cite{Yu2021Spontaneous, lobanov2024generation, ulanov2025quadrature, Jin2025bandgap}. We implement this method for our investigation of the optimal parameters.

\begin{figure*}[ht]
    \centering
    \includegraphics[width=\linewidth]{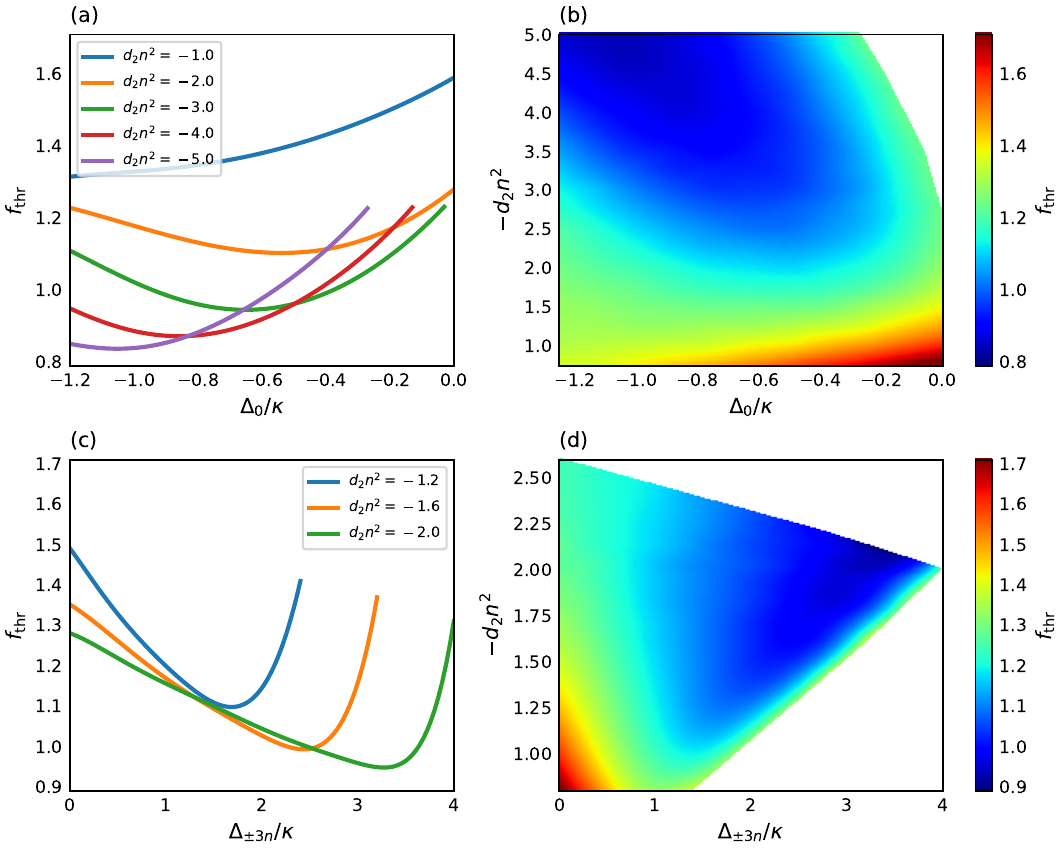}
    \caption{(a) The dependence of the DOPO threshold on the signal mode $\nu = 0$ normalized frequency shift. The lines correspond to the dependence for a different fixed $d_2 n^2$. (b) Colormap for the threshold pump amplitude $f_\text{thr}$ vs normalized shift of the central frequency $\Delta_0$ and the dispersion parameter $-n^2d_2$. The minimum $f=0.835$ is observed at $\Delta_0 = -1.125\kappa$ and $d_2n^2 = -5.0$. White areas correspond to the existence of the modes between pumped ones prior to the generation of the signal mode. Threshold values were calculated for $n=5$. (c) The dependence of the DOPO threshold on the simultaneous modes $\nu = \pm 3n$ frequency shift ($\Delta_{+3n} = \Delta_{-3n}$). The lines correspond to the dependence for a different fixed $d_2 n^2$. (d) Colormap for the threshold pump amplitude $f_\text{thr}$ vs normalized shift of the sideband frequencies (with relative indices $3n$ and $-3n$) $\Delta_{\pm 3n}$ and the dispersion parameter $d_2n^2$. The minimum $f=0.95$ is observed at $\Delta_{\pm3n} = 3.4\kappa$ and $d_2 n^2 = -2.0$. White area corresponds to the existence of the modes between pumped ones prior to the generation of the signal mode. Threshold values were calculated for $n=3$. All quantities are plotted in dimensionless units.}
    \label{fig:fig4}
\end{figure*}

\begin{figure*}[ht]
    \centering
    \includegraphics[width=1\linewidth]{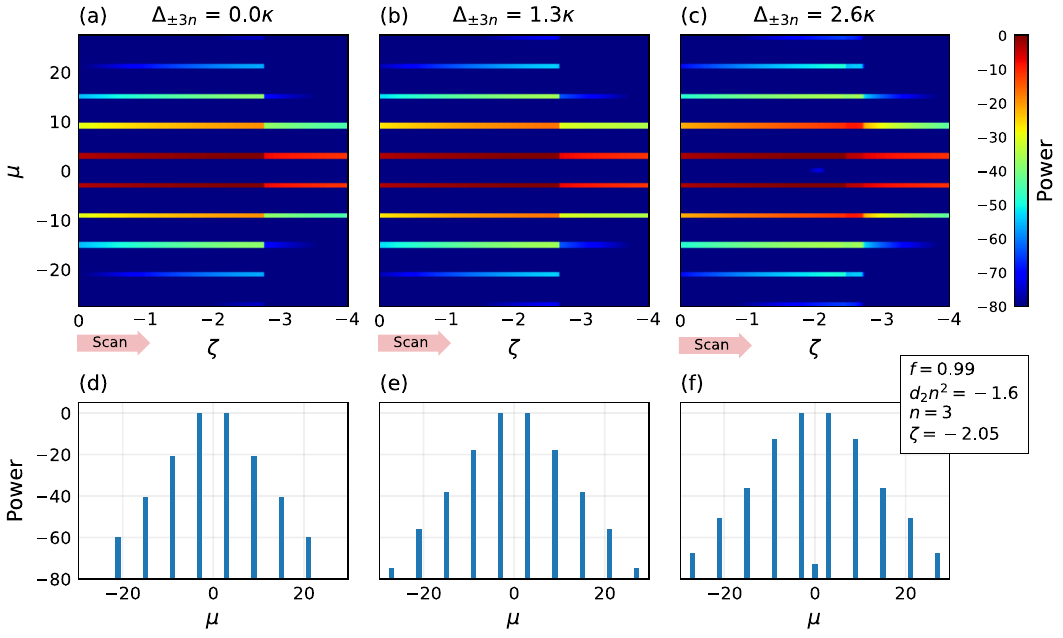}
    \caption{Spectrograms (a-c) for different shift values of the $\pm3n$-th modes for fixed pump amplitude $f=0.99$ and dispersion coefficient $d_2n^2 = -1.6$ and $n=3$. The indices of the shifted modes are $\nu = \pm3n = \pm9$. The energy conversion efficiency from the pumps to the sidebands is increasing with the increment of the shift $\Delta_{\pm 3n}$ (d-f) as shown for the detuning $\zeta = -2.05$ corresponding to the generation of the signal mode (f). Primarily generated sidebands start to participate in the parametric signal generation, thus, lowering required input pumps power. All quantities are plotted in dimensionless units.}
    \label{fig:fig5}
\end{figure*}

Numerically, it can be described by introducing a frequency shift $\Delta_\nu$ of the resonator eigenmodes $\nu$ \cite{Lobanov2015frequency, Yu2021Spontaneous, Xue2015Mode}.
The expression for the frequency of the $\mu$-th mode is:
\begin{equation}\label{eq:dispersionshift}
    \omega_\mu = \omega_0 + \delta _{\mu\nu}\Delta_\nu+ D_1\mu + \frac{1}{2} D_2 \mu^2.
\end{equation}

First, we introduce the frequency shift into the central mode $\nu = 0$. The modified normalized detuning for the central mode, which is introduced in the CMEs, differs from Eq.~(\ref{eq:norm_det}): $\zeta_0^\text{shift} = \zeta_0 - 2 \Delta_0 / \kappa.$ In Fig. \ref{fig:fig4}(a) the dependence of the DOPO threshold on the mode frequency shift for several values $d_2 n^2$ is presented. All the curves are calculated for reasonable $n= 5$. We verified that changing $n$ and $d_2$ accordingly in order to maintain the parameter $d_2n^2$ fixed does not affect the threshold power. 

It is found that introducing a negative shift $\Delta_0 < 0$ of the central mode can significantly reduce the threshold pump power. For each dispersion parameter, there exists an optimal shift value that provides the minimal threshold.
For $d_2 n^2 = -1.0$, the minimum is not pronounced within the range of selected shift values $(|\Delta_0| < 1.2$), and the threshold pump amplitude $f_\text{thr}$ remains well above $1.0$.
As $|d_2| n^2$ increases, a clear and wide minimum appears. The minimum value decreases with increasing $|d_2| n^2$ and drifts to the larger absolute value of $\Delta_0$. For example, we obtained a minimum $f_\text{thr} = 0.835$ for $d_2n^2 = -5.0$, which is more than 50\% lower than for the microring without mode shifting. 

The threshold is calculated over a wide range of parameters $d_2n^2$ and $\Delta_0$. The results are presented in Fig.~\ref{fig:fig4}(b). The white area in the colormap corresponds to the appearance of the non-central modes prior to the DOPO signal. The presence of the excited modes between pumps may decrease the level of squeezing of the output parametric signal. It is worth noting that in the area of low frequency shift ($|\Delta_0|/\kappa < 0.2$) the threshold pump amplitude is relatively high ($f_\text{thr}>1.0$) for any $d_2 n^2$. 

We also study the effects of the symmetrical shifting of the modes $\nu = \pm 3 n$. Such structures are experimentally feasible and are investigated for frequency comb generation in \cite{Jin2025bandgap}. The detunings in the model in this case are $ \zeta_{+3n}^\text{shift} = \zeta_{+3n} - 2 \Delta_{+3n} / \kappa$ and $ \zeta_{-3n}^\text{shift} = \zeta_{-3n} - 2 \Delta_{-3n} / \kappa$. It is shown in Fig.~\ref{fig:fig4}(c) that, for the fixed dispersion parameter, the threshold power slowly decreases with increasing absolute shift value until the minimum for the given dispersion is reached; then we observe an abrupt increase in threshold power. In this case, the threshold power is decreased to $f=0.95$. We also limit the parameter region to the values that allow observation of the signal mode without excited adjacent components. A colormap for the wide range of parameters is shown in Fig.~\ref{fig:fig4}(d).  We confirmed that keeping $d_2n^2$ constant by adjusting $n$ and $d_2$ does not affect the threshold power.

The mechanism underlying this threshold power decrease is illustrated in Fig.~\ref{fig:fig5}: the shift of the sidebands results in the higher energy conversion to other MRR's modes from the pumping. Thus, the parametric signal generation is enhanced via the participation of the secondary modes in the parametric process. 

It is worth noting that the direction of the symmetrical shift of the sidebands $\pm3n$ is different from the shift of the central mode. Also, the shift of the central mode is more effective for large $|d_2|n^2$ values, and it requires smaller absolute shift values compared to the shift of $\pm 3n$ modes.

We verified that the introduced targeted mode shifting does not perturb the bistability of the phase of the parametric signal. This is analyzed via the output signal mode amplitude realizations on the complex plane (described in Appendix~\ref{ap:squeez}).

\section{\label{sec:res}Results and discussion}

We have analytically and numerically derived the threshold pump amplitudes for a DOPO realized in a symmetrically bichromatically pumped single MRR with $\chi^{(3)}$ nonlinearity and normal GVD. Using the three-mode model approximation, we found the optimal parameters and showed analytically the requirement of equal detunings and powers of the pumps. Then we use the CMEs to simulate the amplitudes' evolution in the multimode approach. 
All simulations were conducted in the normalized units, ensuring the relevance of the results across a wide range of photonic platforms.

The threshold parameters obtained from the full multimode model $f=1.24$ differ significantly from those predicted by the approximate three-mode model $f = 1/\sqrt{2}\approx 0.7$, highlighting the importance of considering the full spectrum of mode interactions in the MRR for accurate predictions. Our analysis shows that the minimum of the DOPO pump power threshold occurs at different dispersion parameters in the two models: $d_2n^2 = -1$ for the three-mode model and $d_2n^2 = -2.6$ for the multimode model. Despite these differences, both models show an independence of the DOPO threshold value on the index $n$ of the pumped modes when the product $d_2n^2$ is fixed. Introducing higher-order dispersion approximation into the model reveals that the threshold power increases significantly with increasing frequency interval between the pumps.

We have shown that it is possible to reduce the threshold of the process by introducing shifts of certain modes, specifically, one central mode or two modes with indices $\pm 3n$. It has been shown that, for each dispersion parameter, there is an optimal value of the shift of the corresponding modes. Introducing a frequency shift of $-1.125 \kappa$ in the signal mode at $d_2n^2 = -5.0$, the threshold DOPO pump power can be reduced by more than 50\%, achieving a minimum threshold pump amplitude of $f = 0.835$. That can be of special importance for energy efficiency enhancement of the developed devices and reducing the impact of the thermal effects \cite{Lobanov2023Platicon}.

It was demonstrated experimentally that sophisticated microresonator designs do not significantly degrade the key performance parameters, such as Q-factor and integrated dispersion profile of the system, and provide a perspective platform for nanophotonic devices \cite{Black2022optical, lucas2023tailoring, ulanov2024synthetic, Jin2025bandgap}.
Thus, mode shift approach offers a reasonable approximation for preliminary numerical analysis.
Further research in this direction will help refine the models for practical applications in integrated photonic devices and pave the way for engineering compact (in future, chip-based) and energy-efficient ``turn-key'' devices.
For instance, photonic molecules and photonic crystal microresonators demand distinct modeling approaches that account for secondary effects, including backward or counter-propagating waves and linear or, if required, nonlinear coupling of modes propagating in different directions or in different MRRs \cite{Yang2025Modulational}.

The results we have obtained offer valuable insights into the study and understanding of the four-wave mixing process in a microresonator with bichromatic pumping and advance the practical application of such techniques in the development of various photonic devices.

\begin{acknowledgments}

    The theoretical analysis was supported by Rosatom in the framework of the Roadmap for Quantum computing (Contract No. 868-1.3-15/15-2021 dated October 5).

    The numerical simulations were supported by the Russian Science Foundation (Grant No. 23-42-00111).

    N.~S.~T. developed the theory, with the assistance from G.~V.~G., A.~K.~V. and D.~A.~C.; N.~S.~T. and V.~E.~L. performed the simulations and analyzed the results, with the assistance from A.~V.~M.; N.~S.~T., V.~E.~L., A.~E.~S. and D.~A.~C. wrote the manuscript, with input from others;
    I.~A.~B., D.~A.~C. and V.~E.~L. supervised the project.
    All authors discussed the results and reviewed the manuscript.
    
    The authors declare no competing interests.
\end{acknowledgments}

\section*{Data Availability}
The data that support the findings of this study are available from the authors upon reasonable request.

\begin{figure}[t]
    \centering\includegraphics[width=\linewidth]{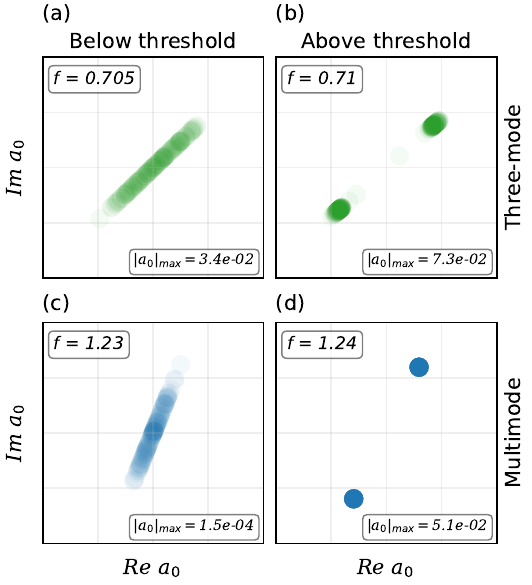}
    \caption{Realizations of complex amplitude of the signal mode $a_0$ for different regimes in the three-mode model with the optimal parameters $d_2n^2 = -1.0$ (a,b) and in the multimode model with the optimal parameters $d_2n^2 = -2.4$ (c,d). All quantities are plotted in dimensionless units.}
    \label{fig:fig6}
\end{figure}
\begin{figure}[t]
    \centering\includegraphics[width=\linewidth]{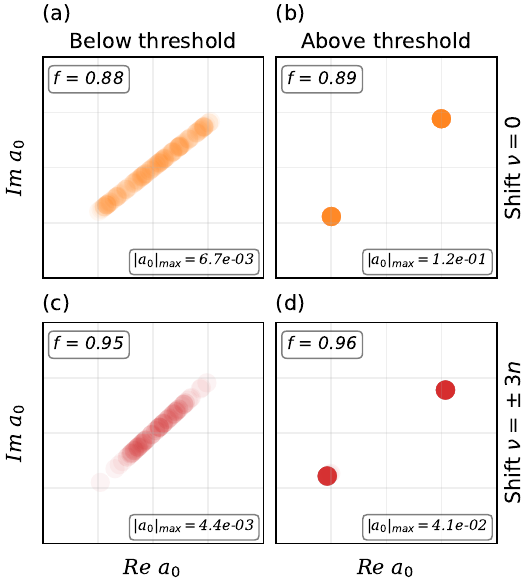}
    \caption{Realizations of complex amplitude of the signal mode $a_0$ for different regimes in the model with the shift of the central mode for parameters $d_2n^2 = -4.0$ and $\Delta_0 = -0.75\kappa$ (a,b) and in the model with the shifts of the sidebands for parameters $d_2n^2 = -2.0$ and $\Delta_{\pm 3n} = 3.4\kappa$ (c,d). All quantities are plotted in dimensionless units.}
    \label{fig:fig7}
\end{figure}

\appendix

\section{\label{ap:CMEs}Coupled mode equations}

The field dynamics inside the MRR can be described using the slow varying amplitude approximation at the signal resonance frequency $\omega_0$. The CMEs for the modes of the MRR are given by:
    \begin{multline}\label{eq:CME_init}
    \frac{\partial A_\mu}{\partial t} = -\frac{\kappa}{2} A_\mu + \delta_{+n,\mu} \sqrt{\kappa_\text{ext} \frac{P_{\text{in}, +n}}{\hbar \omega_{p+}}}e^{-i\left( \omega_{p+} - \omega_{+n} \right)t} + \\
    \delta_{-n,\mu} \sqrt{\kappa_\text{ext} \frac{P_{\text{in}, -n}}{\hbar \omega_{p-}}}e^{-i\left( \omega_{p-} - \omega_{-n} \right)t} + \\
    ig \sum_{\mu'\mu''\mu'''}A_{\mu'}A_{\mu''}A_{\mu'''}^* e^{\left( \omega_{\mu'} + \omega_{\mu''} -\omega_{\mu'''} -\omega_{\mu}  \right)},
\end{multline}
where $A_\mu$ is the amplitude of the MRR's eigenfrequency, normalized such that $|A_\mu|^2$ corresponds to the number of photons in that mode, $\kappa_\text{ext}$ is the coupling rate and $\kappa = \kappa_\text{int} + \kappa_\text{ext}$ is the full resonator decay rate with $\kappa_\text{int}$ denoting the intrinsic loss of the MRR. The nonlinear coupling coefficient $g$ is given by:
\begin{equation}
    g = \frac{\hbar \omega_0^2 c n_2}{n_0^2 V_\text{eff}},
\end{equation}
where $n_0$ is the refractive index, $n_2$ is the nonlinear refractive index, $V_\text{eff}$ is the effective mode volume.

The pumps are introduced in Eq.~(\ref{eq:CME_init}) via pump powers $P_{\text{in}, \pm}$ and pump frequencies $\omega_{p\pm}$, which are detuned from the MRR's eigenfrequencies $\omega_{\pm n}$, respectively.

We normalize these equations as follows:
\begin{equation}\label{eq:normalization}
    a_\mu = \sqrt{\frac{2g}{\kappa}} A_\mu e^{-\left(\omega_\mu - \frac{\omega_{p+} + \omega_{p-}}{2}- \mu \frac{\omega_{p+} - \omega_{p-}}{2 n}  \right)t },
\end{equation}
where $a_\mu$ is the normalized mode amplitude. This normalization results in the time-independent terms and accounts for the phase shift due to deviations from an equidistant frequency grid. With bichromatic pumping, the effective FSR of the system is $(\omega_{p+} - \omega_{p-})/(2n)$.

From Eq.~(\ref{eq:normalization}) other values are normalized as follows: $\tau = t \kappa / 2$ is the slow time, the pump amplitudes $f_{\pm n} = \sqrt{\frac{8 \kappa_\text{ext} g}{\kappa^3}} \sqrt{\frac{P_{\text{in}, \pm }}{\hbar \omega_{0 }}}$. 
Formally, it is necessary to normalize the pump amplitudes by their respective frequencies; however, we assume that $\omega_0$ and $\omega_{p \pm}$ are close enough that their difference can be neglected. 
The normalized second-order coefficient is given by $d_2 = 2D_2 /\kappa$ and detuning [see Eq.~(\ref{eq:norm_det})]:
\begin{equation}\label{eq:norm_det_app}
    \zeta_\mu = \frac{\zeta_{+n} + \zeta_{-n}}{2} + \frac{\zeta_{+n} - \zeta_{-n}}{2n}\mu + \frac{d_2}{2} (n^2 - \mu^2),
\end{equation}
where $\zeta_{\pm n} = 2 \left( \omega_{p\pm} - \omega_{\pm n} \right) / \kappa$ denote the normalized pump detunings from the corresponding MRR eigenmode.

\section{\label{ap:squeez}Squeezing}

In dual-pumped systems operating below the DOPO threshold regime, a squeezed vacuum state is generated in the signal mode. If the pump power is above the threshold, the signal field bifurcates to two possible states with phases offset by $\pi$. To verify the obtained threshold parameters, we perform numerical simulations of the experimental process, collecting statistical data from simulations with various initial vacuum noise seed fields. 

Firstly, we examine both the three-mode and multimode models (see Fig.~\ref{fig:fig6}). In both models, the generated signal field exhibits quadrature squeezing for input powers below the threshold values. Over one hundred realizations at low tuning speed were made to calculate the data for each panel of Fig. ~\ref{fig:fig6}. In the above-threshold regime, following a non-equilibrium transition, the generated signal stabilizes in one of two possible states with a phase offset of $\pi$.

Secondly, we perform the same procedure using the models with shifts. The results (Fig.~\ref{fig:fig7}) coincide with the threshold definition.

It is important to note that these results reflect the squeezing \textit{inside} the MRR. The output signal, however, will be affected due to coupling losses. As the theoretical results are expressed in normalized units, the threshold power value $f_\text{thr}$ does not change, but the coupling is taken into account in its normalization, as detailed in Appendix~\ref{ap:CMEs}. Consequently, the output squeezing factor is highly dependent on the input power in real units. To achieve higher squeezing, it is necessary to increase the coupling coefficient and, hence, the input power.

\bibliography{bibliography}

\end{document}